\documentclass{jpsj-suppl}
\usepackage{times}
\usepackage{color}
\def\be{\begin{equation}}
\def\ee{\end{equation}}
\def\bea{\begin{eqnarray}}
\def\eea{\end{eqnarray}}
\title{Berezinskii-Kosterlitz-Thouless-type transitions in $d\,{=}\,2$ quantum O(2) and O(2)$\times$O(2) nonlinear sigma models}

\author{C.A. Hooley$^{1}$, S.T. Carr$^{2}$, J.M. Fellows$^{3}$, and J. Schmalian$^{4}$} 
\inst{$^{1}$Scottish Universities Physics Alliance, School of Physics and Astronomy, University of St Andrews, North Haugh, St Andrews, Fife KY16 9SS, UK \\
$^{2}$School of Physical Sciences,
University of Kent, Canterbury CT2 7NH, UK \\
$^{3}$Department of Physics, University of Warwick, Coventry CV4 7AL, UK \\
$^{4}$Institut f{\"u}r Theorie der Kondensierten Materie and DFG Center for Functional Nanostructures,
Karlsruher Institut f{\"u}r Technologie, 76128 Karlsruhe, Germany} 
\abst{We discuss the $d\,{=}\,2$ quantum O(2)$\times$O(2) nonlinear sigma model as a low-energy theory of phase reconstruction near a quantum critical point.  We first examine the evolution of the Berezinskii-Kosterlitz-Thouless (BKT) transition as the quantum limit is approached in the usual O(2) nonlinear sigma model.  Then we go on to review results on the ground-state phase diagram of the O(2)$\times$O(2) nonlinear sigma model, and on the behaviour of the O(2)$\times$O($M$) nonlinear sigma model with $M>2$ in the classical limit.  Finally, we present a conjectured finite-temperature phase diagram for the quantum version of the latter model in the O(2)$\times$O(2) case.  The nature of the finite-temperature BKT-like transitions in the phase diagram is discussed, and avenues for further calculation are identified.}

\kword{Phase reconstruction, quantum critical points, Berezinskii-Kosterlitz-Thouless transition}

\begin{document}
\maketitle
\section{Motivation}
One of the key questions of modern condensed matter physics is that of phase reconstruction near quantum critical points.  Examples include the development of superconductivity around the quantum critical points in heavy-fermion intermetallics \cite{hfsupercond}, the occurrence of various types of textured magnetic order in itinerant ferro- or helimagnets \cite{magtext,kruger2012}, and the undiagnosed phase at low temperature and high magnetic field in Sr$_3$Ru$_2$O$_7$ \cite{sr327}.  Simple model systems for studying such phase coexistence are, however, hard to come by.  In particular, in all of the abovementioned examples the situation is complicated by the fact that the background is metallic, which means that there are always low-energy excitations (the particle-hole modes) in addition to the ones that `drive' the transition.

An important recent development in this field is the observation by Jaefari, Lal, and Fradkin \cite{jaefari2010}, based on earlier work by Calabrese {\it et al.\/} \cite{calabrese2003}, that the $d\,{=}\,2$ quantum O(2)$\times$O(2) nonlinear sigma model also exhibits a coexistence phase near its O(4) high-symmetry point.  Since there is no metallic background in this problem, it may prove a more tractable starting point for the study of phase coexistence phenomena.  In addition, the fact that the model is in $d\,{=}\,2$ allows the possibility of some rather exotic Berezinskii-Kosterlitz-Thouless (BKT) physics \cite{berezinskii,kt} in its finite-temperature phase diagram.  The O(2)$\times$O(2) nonlinear sigma model has already been proposed as a model of phase competition in a system of dipolar bosons in a quasi-one-dimensional optical lattice \cite{fellows2011}; in that case, the coexistence phase is a supersolid.

In this paper, we begin by pointing out some underemphasised issues in the finite-temperature phase diagram of the $d\,{=}\,2$ quantum O(2) nonlinear sigma model.  We then review known properties of the $d\,{=}\,2$ quantum O(2)$\times$O(2) nonlinear sigma model at zero temperature, before proceeding to analyse its finite-temperature phase diagram near the phase-coexistence region.  We summarise our recent calculation \cite{ourprl} of the behaviour of the O(2)$\times$O($M$) nonlinear sigma model near the high-symmetry point in the classical limit, and speculate on the fate of the quantum BKT transition in the O(2)$\times$O(2) model as the high-symmetry point is approached.
\section{The quantum O(2) nonlinear sigma model in two spatial dimensions}
\subsection{Definition of the model}
The action for the quantum O(2) nonlinear sigma model in two spatial dimensions is
\be
S = \frac{g}{2} \int\limits_0^\beta d\tau \int d^2 x \left[ \left( \partial_\tau {\bf n} \right)^2 + \left( \partial_x {\bf n} \right)^2 + \left( \partial_y {\bf n} \right)^2 \right], \label{o2nlsm}
\ee
where ${\bf n}$ is a two-component vector, $\beta \equiv 1/T$ is the inverse temperature, and $g$ is a (renormalised low-energy) stiffness parameter.  We work in units where $\hbar=k_B=1$.  The nonlinearity is provided by a unit-length constraint on the vector ${\bf n}$, viz.\ that ${\bf n}^2 = 1$.
\subsection{High-temperature (classical) behaviour}
At high temperatures (i.e.\ as $\beta \to 0$) the model (\ref{o2nlsm}) behaves classically.  This is because the field ${\bf n}({\bf x},\tau)$ is constrained to be periodic in the $\tau$-direction; thus, as the $\tau$-interval $(0,\beta)$ shrinks, the non-uniform Fourier components in ${\bf n}({\bf x},\tau)$ acquire increasingly large actions and become physically irrelevant.  The action in the high-temperature limit is therefore given by
\be
S_{\rm cl} = \frac{g}{2T} \int d^2 x \left[ \left( \partial_x {\bf n} \right)^2 + \left( \partial_y {\bf n} \right)^2 \right]. \label{o2nlsmclass}
\ee
The unit-length constraint is automatically obeyed if we use the parameterisation ${\bf n} = \left( \cos\theta,\sin\theta \right)$, in terms of which the classical action becomes
\be
S_{\rm cl} = \frac{g}{2T} \int d^2 x \left[ \left( \partial_x \theta \right)^2 + \left( \partial_y \theta \right)^2 \right]. \label{o2nlsmclasstheta}
\ee
Hence the propagator of the $\theta$-field goes like $1/k^2$, and it would seem to follow that the correlation functions in the model exhibit algebraic decay at all temperatures.

\begin{figure}
\begin{center}
\includegraphics[width=14cm]{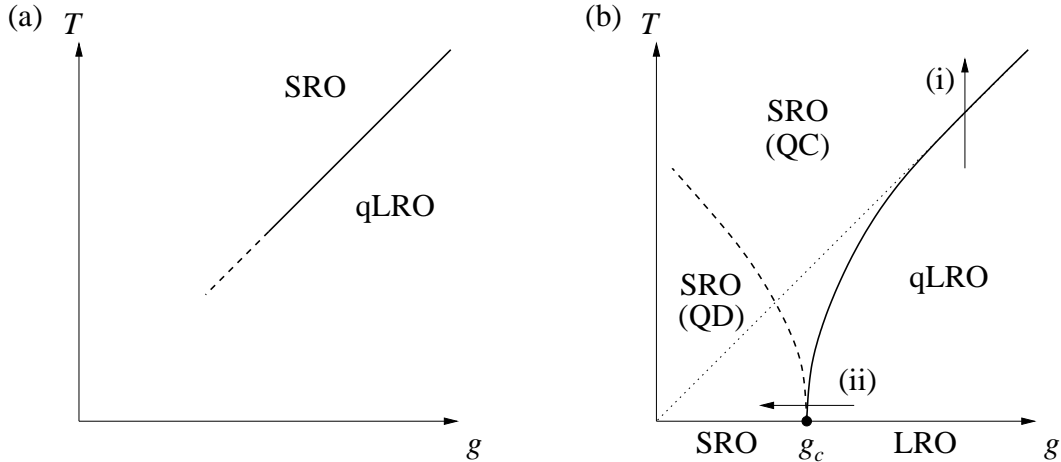}
\end{center}
\caption{(a) The high-temperature (= classical) part of the phase diagram of the quantum O(2) nonlinear sigma model.  The line represents a Berenzinskii-Kosterlitz-Thouless (BKT) transition at $T_c \approx 0.89 \pi g$.  The labels `qLRO' and `SRO' mean respectively `quasi-long-range-ordered' (i.e.\ algebraically decaying correlations) and `short-range-ordered'. (b) The full phase diagram of the quantum O(2) nonlinear sigma model, obtained by joining up the classical limit with the known $T=0$ behaviour.  The critical stiffness at which the system disorders at $T=0$ is indicated by $g_c$.  The labels `LRO', `QC', and `QD' mean respectively `long-range-ordered', `quantum critical', and `quantum disordered'.  The natures of these regions, including the distinction between quantum-critical and quantum-disordered behaviour, are discussed in the text.  Labels below the horizontal axis apply to the $T=0$ state of the model in that range of stiffness.}
\label{o2nlsmpd}
\end{figure}

However, it is clear on physical grounds that at high temperatures the model should exhibit short-range order, with a correlation length tending to zero as $T \to \infty$.  The resolution of this apparent paradox was provided by Berezinskii \cite{berezinskii} and Kosterlitz and Thouless \cite{kt}:\ there is a finite-temperature topological transition in the model associated with the unbinding of vortex--anti-vortex pairs in the field ${\bf n}({\bf x})$.  This is known as the BKT transition; it occurs at a temperature $T_{\rm BKT} \approx 0.89 \pi g$ \cite{kt}.
Thus the classical part of the phase diagram of (\ref{o2nlsmclass}) is as shown in Fig.~\ref{o2nlsmpd}(a).
\subsection{Low-temperature behaviour}
What happens to this transition line as we proceed to lower temperatures, thereby invalidating the classical approximation?  The answer may be inferred by considering another limit in which the behaviour of the model (\ref{o2nlsm}) is known, viz.\ the $T=0$ limit.  In this limit, the $\tau$-interval becomes infinite, and then (because of the symmetric nature of the Lagrangian density) $\tau$ behaves as another spatial co-ordinate.  Hence the phase diagram of the $T=0$ quantum model as a function of inverse stiffness is the same as that of the {\it three}-dimensional classical model as a function of temperature.  This is the well-known quantum-classical correspondence \cite{qclass}.  Since the $d\,{=}\,3$ classical O(2) nonlinear sigma model has a conventional spin-wave-driven transition from long-range order to short-range order at a critical temperature $T_c$ \cite{campostrini2001}, it follows that the $d\,{=}\,2$ zero-temperature quantum O(2) nonlinear sigma model has one at a critical stiffness $g_c$.  This expectation is borne out by recent numerical work \cite{langfeld2013}.

On the (apparently plausible) assumption that one can have quasi-long-range-order (i.e.\ algebraically decaying correlations) at $T \ne 0$ only for stiffnesses at which there is true long-range-order in the ground state, we must extend the BKT line to join up with the quantum critical point at $g=g_c$.  This yields a phase diagram reminiscent of that provided by Chakravarty, Halperin, and Nelson \cite{chn} for the $d\,{=}\,2$ O(3) nonlinear sigma model, but with their short-range-ordered (renormalised classical) region replaced by a quasi-long-range-ordered phase.  We emphasise that for $g>g_c$ the ground state of the O(2) model has true long-range order, which is separated by a first-order transition from the quasi-long-range order pertaining at finite temperatures.  In the O(2) case, just as in the O(3) case, we expect a crossover for $g<g_c$ from a quantum disordered to a quantum critical region as the temperature is increased.  The distinction between these regions is the temperature-dependence of the correlation length, $\xi$ \cite{chn}:\ in the quantum disordered region we expect $\xi \sim \xi_0$, a temperature-independent constant; in the quantum critical region, by contrast, $\xi \sim 1/T$.
\subsection{A comment on the r{\^o}le of vortices}
This poses an interesting question of interpretation.  If we drive the system across the qLRO--SRO phase transition following trajectory (i) in Fig.~\ref{o2nlsmpd}(b), the BKT analysis applies, and thus we describe the transition in terms of the unbinding of vortex--anti-vortex pairs.  If, however, we drive the system across the qLRO--SRO transition at very low temperatures, following trajectory (ii), the proximity to the quantum critical point suggests that spin waves must be the main driving mechanism.

\begin{figure}
\begin{center}
\leavevmode
\includegraphics[width=13.5cm]{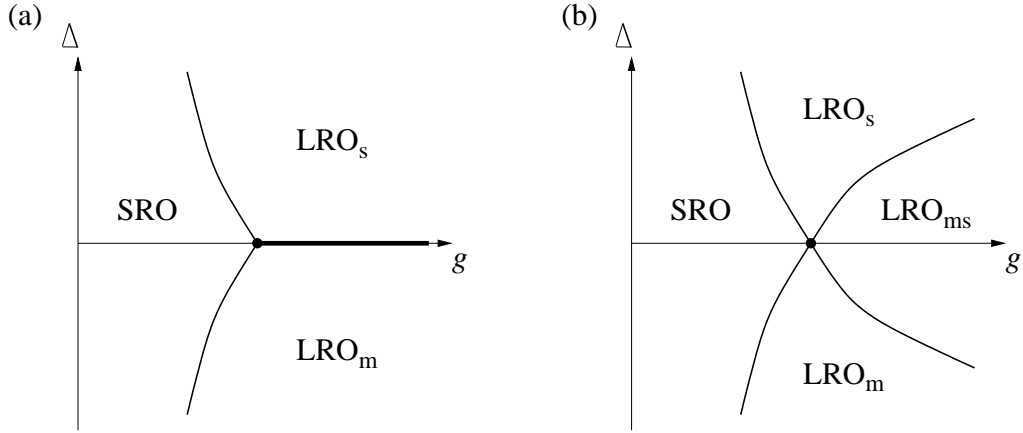}
\end{center}
\caption{(a) The $T=0$ phase diagram of the $d\,{=}\,2$ quantum O(2)$\times$O(2) nonlinear sigma model as obtained in Refs.~\cite{prejlf}.  The labels `LRO$_{\rm s}$', `LRO$_{\rm m}$', and `SRO' signify respectively: long-range order with $\langle {\bf n} \rangle$ lying in the ${\bf s}$-subspace; long-range order with $\langle {\bf n} \rangle$ lying in the ${\bf m}$-subspace; and short-range order.  The thick line represents a first-order transition; all other transitions shown are continuous.  (b) The $T=0$ phase diagram according to Jaefari, Lal, and Fradkin \cite{jaefari2010} (based on earlier work by Calabrese {\it et al.}\ \cite{calabrese2003}).  The new phase `LRO$_{\rm ms}$' is a coexistence phase, in which both $\langle {\bf m} \rangle$ and $\langle {\bf s} \rangle$ are non-zero.  Put another way, the vector $\langle {\bf n} \rangle$ in this phase does not lie purely in either of the large-$\vert \Delta \vert$ subspaces.}
\label{o2o2nlsmt0}
\end{figure}
One obvious possibility is that the appropriate description depends on proximity to the transition in the usual fashion \cite{qclass}:\ there is always a `window of classicality' near the transition, and in this window the BKT vortex picture applies; but this window becomes narrower and narrower as the quantum critical point is approached.  However, it is also interesting to note recent work by Holzmann {\it et al.\/} \cite{holzmann}, in which a description of the finite-temperature transition is given in which vortices are apparently absent.  In our opinion, this question deserves further investigation.
\section{Competition with another ordered phase at low temperatures:\ the quantum O(2)$\times$O(2) nonlinear sigma model}
Let us now turn to the question of what happens to the quasi-long-range-ordered phase, and its associated BKT transition, when the system is brought close to the boundary of a different ordered phase.  To this end, we consider the $d\,{=}\,2$ quantum O(2)$\times$O(2) nonlinear sigma model.
\subsection{Definition of the model}
The action of this nonlinear sigma model is
\be
S = \frac{g}{2} \int\limits_0^\beta d\tau \int d^2 x \left[ \left( \partial_\tau {\bf n} \right)^2 + \left( \partial_\mu {\bf n} \right)^2 + \frac{\Delta}{a^2} {\bf m}^2 \right]. \label{o2o2nlsm}
\ee
Here $g$ is again a (renormalised low-energy) stiffness parameter, and $a$ is a short-distance cut-off.  The vector ${\bf n}$ now has {\it four} components: ${\bf n} = \left( {\bf s},{\bf m} \right)$,
where ${\bf s}$ and ${\bf m}$ have two components each.  The unit-length constraint, however, is applied to ${\bf n}$ as a whole, so that ${\bf s}^2 + {\bf m}^2 = 1$.
This allows the redistribution of weight between the ${\bf s}$- and ${\bf m}$-sectors as the tuning parameter $\Delta$ is changed.  In particular, for $\Delta \to +\infty$, all finite-action configurations will have ${\bf n}$ lying entirely in the ${\bf s}$-subspace, while for $\Delta \to -\infty$ they will have ${\bf n}$ lying entirely in the ${\bf m}$-subspace.  Note also that when $\Delta=0$ the symmetry of the model is enhanced from O(2)$\times$O(2) to O(4).  At this point, in accordance with the Mermin-Wagner theorem \cite{merminwagner}, there can be no finite-temperature phase transitions, even of the BKT type.  However, even though any BKT lines must go to $T=0$ at $\Delta=0$ for this reason, it does not necessarily follow that $(\Delta,T)=(0,0)$ is a quantum critical point of the model.
\begin{figure}
\begin{center}
\includegraphics[width=14cm]{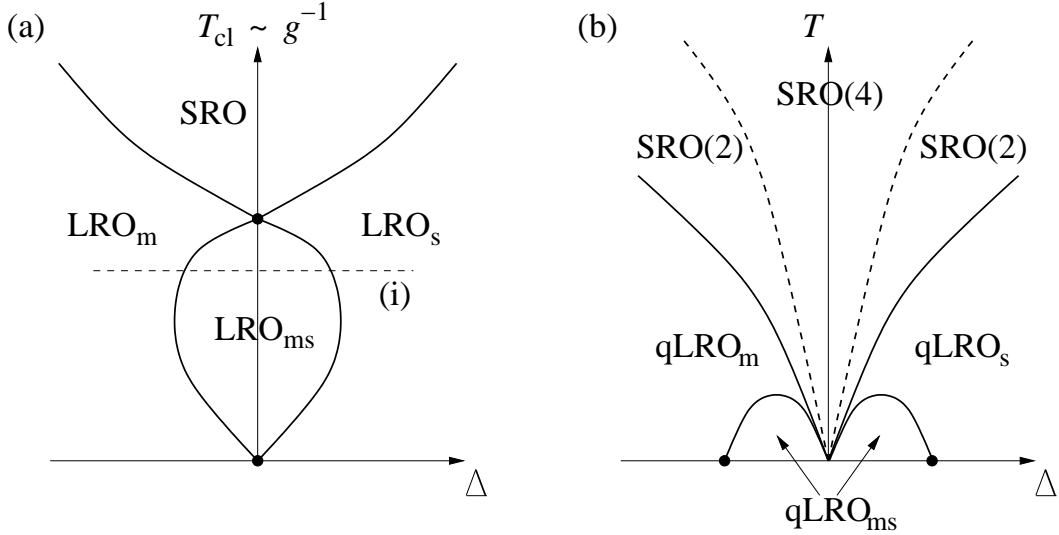}
\end{center}
\caption{(a) A likely phase diagram of the $d\,{=}\,3$ classical O(2)$\times$O(2) nonlinear sigma model.  According to the quantum-classical correspondence \cite{qclass}, the vertical axis may be thought of either as the temperature of the $d=3$ classical model ($T_{\rm cl}$) or as the inverse stiffness parameter of the $d=2$ quantum model ($g^{-1}$).  Notice that the coexistence phase, being in the $d=3$ representation a thermal order-by-disorder effect, must vanish as $T_{\rm cl} \to 0$.  This corresponds, in the $d\,{=}\,2$ quantum model, to the high-stiffness limit.  (b) A likely phase diagram of the $d\,{=}\,2$ quantum O(2)$\times$O(2) nonlinear sigma model, for a stiffness indicated by the dashed line (i) in panel (a).  The solid lines are BKT-type transitions; the dashed lines are crossovers.  The region labelled `qLRO$_{\rm m}$' has quasi-long-range order (i.e.\ algebraically decaying correlations) in the ${\bf m}$-subspace; that labelled `qLRO$_{\rm s}$' has quasi-long-range order in the ${\bf s}$-subspace; and that labelled `qLRO$_{\rm ms}$' is the coexistence phase, which has quasi-long-range order in both subspaces.  The regions `SRO(2)' are short-range-ordered regions in which the fluctuations are dominated by vortices in the appropriate two-dimensional subspace; the region `SRO(4)' is the renormalised classical region emanating from the $\Delta=T=0$ high-symmetry point.  In this region the fluctuations are spin-wave-like.  We emphasise that the point $\Delta=T=0$ is {\it not\/} a quantum critical point, even though (for symmetry reasons) the BKT transition temperatures must all vanish as $\Delta \to 0$.}
\label{thermalobdo}
\end{figure}
\subsection{Zero-temperature behaviour}
It was believed for some time \cite{prejlf} that the $T=0$ phase diagram of the model (\ref{o2o2nlsm}) exhibited only three phases:\ LRO$_{\rm s}$ (long-range order with $\langle {\bf n} \rangle$ lying in the ${\bf s}$-subspace); LRO$_{\rm m}$ (long-range order with $\langle {\bf n} \rangle$ lying in the ${\bf m}$-subspace); and SRO (short-range order).  This state of affairs is depicted schematically in Fig.~\ref{o2o2nlsmt0}(a).

However, it was argued by Jaefari, Lal, and Fradkin in 2010 \cite{jaefari2010}, based on earlier work by Calabrese {\it et al.}\ \cite{calabrese2003}, that in fact there is a fourth phase in the diagram:\ a long-range-ordered phase in which the vector $\langle {\bf n} \rangle$ does not lie purely in either of the large-$\vert \Delta \vert$ subspaces.  We emphasise that their chief argument is based on a Landau expansion, and therefore can predict only the behaviour in the vicinity of the tetracritical point.

The physical origin of such a coexistence phase must be some kind of quantum order-by-disorder effect; such mechanisms have already been discussed by other authors \cite{kruger2012} in the context of phase reconstruction near certain metallic quantum critical points.  It is desirable to understand this effect better.  To this end, we may exploit the quantum-to-classical correspondence mentioned above:\ the $(\Delta,g)$ phase diagram of the $d\,{=}\,2$ quantum model must be the same as the $(\Delta,T^{-1})$ phase diagram of the $d\,{=}\,3$ classical model.  In the latter picture, the coexistence phase must be created by a thermal order-by-disorder mechanism; a likely phase diagram is shown in Fig.~\ref{thermalobdo}(a).  Notice that, in the classical model, the coexistence phase must vanish at $T=0$ since there can be no thermal effects there.
An advantage of the $d\,{=}\,3$ classical approach is that it can be studied by classical Monte Carlo methods.  Preliminary studies in this direction have already been undertaken \cite{staffini2013}, but reasonably large systems are likely to be required for definitive results.
\subsection{Finite-temperature behaviour}
The finite-temperature behaviour of the model (\ref{o2o2nlsm}) is complicated, because of the existence of two types of vortex, one associated with each large-$\vert \Delta \vert$ subspace.  The interplay of these vortices is a difficult topic which is still not wholly resolved.  However, if we generalise the model to O(2)$\times$O($M$) and choose $M>2$, things become simpler.  In that case we have an O(2)-type long-range order (which supports vortices) at large positive $\Delta$, while we have an O($M$)-type long-range order (which supports spin waves) at large negative $\Delta$.  In this case, as shown in some of our recent work \cite{ourprl}, the approach to the high-symmetry point in the classical limit can be described.

The main effect of the approach to $\Delta=0$ is an `eating away' of the vortices in the O(2) sector from the interior, in a phenomenon somewhat reminiscent of skyrmion formation --- except that instead of having one out-of-plane direction in which to point, the vector ${\bf n}$ has $M$ such directions at its disposal.  As a result, there is an additional crossover line in the finite-temperature phase diagram compared to that sketched by Jaefari {\it et al.}\ \cite{jaefari2010}, across which the nature of the fluctuations changes from O(2) (vortex-like) to O($M$+2) (spin-wave-like).

We might conjecture that similar physics will apply in the O(2)$\times$O(2) model, though because this scenario now involves `vortices within vortices' it is a technically more challenging situation to address.  Assuming, however, that the crossover line still exists in this case, we would obtain a phase diagram for the quantum O(2)$\times$O(2) model somewhat like that shown in Fig.~\ref{thermalobdo}(b).  The detailed nature of the BKT-like transitions bounding the `LRO$_{\rm ms}$' regions would merit further research.
\section{Summary}
In this paper, we have presented the following:\ a conjectured phase diagram of the quantum O(2) nonlinear sigma model, Fig.~\ref{o2nlsmpd}(b); a comment on the r{\^o}le of vortices in its Berezinskii-Kosterlitz-Thouless transition as the temperature is lowered into the quantum regime; a conjectured phase diagram of the quantum O(2)$\times$O(2) nonlinear sigma model at $T=0$, Fig.~\ref{thermalobdo}(a); and a summary of our recent results, plus some further conjectures, about that model's finite-temperature phase diagram.  We have also indicated several possible directions for further work on this problem.

\end{document}